\documentclass[journal=nalefd,manuscript=letter]{achemso}

\usepackage{chemformula} 
\usepackage[T1]{fontenc} 
\usepackage{amssymb}

\usepackage{color}

\definecolor{Orange}{rgb}{0.95,0.46,0.17}
\usepackage{soul}

\author{Francesca Telesio}
\affiliation{NEST, Istituto Nanoscienze-CNR and Scuola Normale Superiore, Piazza San Silvestro 12, 56127 Pisa, Italy}

\author{Matteo Carrega}
\affiliation{CNR-SPIN, Via Dodecaneso 33, 16146, Genova, Italy}

\author{Giulio Cappelli}
\affiliation{NEST, Istituto Nanoscienze-CNR and Scuola Normale Superiore, Piazza San Silvestro 12, 56127 Pisa, Italy}

\author{Andrea Iorio}
\affiliation{NEST, Istituto Nanoscienze-CNR and Scuola Normale Superiore, Piazza San Silvestro 12, 56127 Pisa, Italy}

\author{Alessandro Crippa}
\affiliation{NEST, Istituto Nanoscienze-CNR and Scuola Normale Superiore, Piazza San Silvestro 12, 56127 Pisa, Italy}

\author{Elia Strambini}
\affiliation{NEST, Istituto Nanoscienze-CNR and Scuola Normale Superiore, Piazza San Silvestro 12, 56127 Pisa, Italy}

\author{Francesco Giazotto}
\affiliation{NEST, Istituto Nanoscienze-CNR and Scuola Normale Superiore, Piazza San Silvestro 12, 56127 Pisa, Italy}

\author{Manuel Serrano--Ruiz}
\affiliation{CNR-ICCOM, Via Madonna del Piano 10, 50019 Sesto Fiorentino, Italy}

\author{Maurizio Peruzzini}
\affiliation{CNR-ICCOM, Via Madonna del Piano 10, 50019 Sesto Fiorentino, Italy}

\author{Stefan Heun}
\email{stefan.heun@nano.cnr.it}
\affiliation{NEST, Istituto Nanoscienze-CNR and Scuola Normale Superiore, Piazza San Silvestro 12, 56127 Pisa, Italy}

\title{Evidence of Josephson coupling in a few-layer black phosphorus planar Josephson junction}

\begin{document}

\begin{tocentry}
\includegraphics{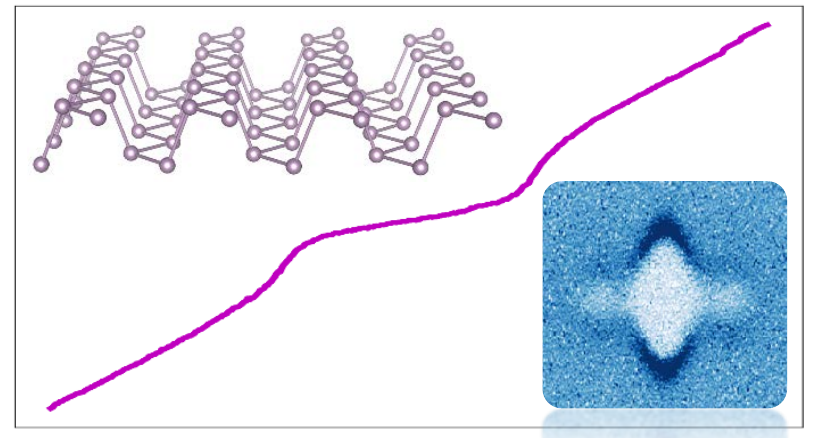}
\end{tocentry}

\begin{abstract}
Setting up strong Josephson coupling in van der Waals materials in close proximity to superconductors offers several opportunities both to inspect fundamental physics and to develop novel cryogenic quantum technologies. Here we show evidence of Josephson coupling in a planar few-layer black Phosphorus junction. The planar geometry allows us to probe the junction behavior by means of external gates, at different carrier concentrations. Clear signatures of Josephson coupling are demonstrated by measuring supercurrent flow through the junction at milli Kelvin temperatures. Manifestation of Fraunhofer pattern with a transverse magnetic field is also reported, confirming the Josephson coupling. These findings represent the first evidence of proximity Josephson coupling in a planar junction based on a van der Waals material beyond graphene and open the way to new studies, exploiting the peculiar properties of exfoliated black phosphorus thin flakes.
\end{abstract}


To date, Josephson junction (JJ) devices find a wide range of applications, such as highly sensitive magnetometers \cite{Krey1999}, infrared sensors \cite{Grimes1968,Osterman1991,Walsh2021}, single-photon detectors \cite{Walsh2021}, and superconducting quantum circuits for quantum information purposes \cite{Kockum2019}. Moreover, hybrid JJs formed with semiconductors placed in between two superconducting contacts have attracted great interest \cite{Beenakker1991,Marsh1994,Heida1998,Schaepers2001,Doh2005,Larsen2015,Tiira2017,Strambini2020,Aggarwal2021,Moehle2021}, due to the possibility of tuning their superconducting properties by external electrostatic gates. Among all, high-quality two dimensional electron gases (2DEGs) have been inspected in so-called planar JJs, exploiting peculiar features such as long ballistic transport and conductance quantization in combination with proximity-induced superconducting correlations \cite{Chrestin1994,Mohammadkhani2008,Amado2013,Shabani2016,Fornieri2019,Guiducci2019,Guiducci2019a,Moehle2021}.

In the study of 2D materials, recent advances in mechanical exfoliation techniques allowed for the fabrication of devices with thickness ranging from a single 2D layer, like graphene and its van der Waals relatives, to multilayer systems with different and tunable properties \cite{Geim2007,CastroNeto2009,Novoselov2012,Geim2013,Huang2017,Jariwala2017,Liu2019,Liu2014,Ling2015,CastellanosGomez2015,Peruzzini2019}. Planar junctions based on novel 2D exfoliated materials constitute a very interesting and intriguing platform for several reasons. To start with, such devices offer the great opportunity to study fundamental aspects of distinct physics, such as Dirac-like relativistic behavior in combination with superconductivity \cite{GrapheneJJs, Calado2015, Lee2018, BenShalom2016}, and are a promising platform for advanced sensing and detection down to the single photon quantum in a wide spectral range\cite{Walsh2021}. Importantly, the superconducting contacts can be placed laterally, retaining the surface of the device free, allowing for the spatially resolved investigation of correlations and transport properties, providing also the possibility to add top gates. This has triggered great experimental efforts to realize hybrid JJs with 2D flakes, and led in 2007 to the first fabrication and characterization of a graphene-based JJ device \cite{GrapheneJJs}, demonstrating a bipolar tuning of supercurrent amplitude with well developed Josephson coupling.

Since then, a vast variety of graphene-superconductor hybrid structures have been explored (see Ref.~\cite{Lee2018} and references therein), with an improved quality of both the graphene samples and the transparency of the interfaces. This allowed to investigate proximity-induced superconductivity in graphene junctions from the diffusive to the ballistic regime. Interestingly, the coexistence of superconductivity and quantum Hall regime has been reported recently in this platform \cite{Amet2016, Park2017, Lee2017, Guel2021}, promoting graphene-based JJs as a good candidate to investigate the  emergence of new topological states of matter \cite{Stern2013,Carrega2021}.

Graphene is only one example of the large family of van der Waals materials that can be exfoliated down to the monolayer thickness \cite{Geim2013,Huang2017,Jariwala2017,Liu2019,Liu2014,Ling2015,CastellanosGomez2015,Peruzzini2019}. However, up to date, clear signatures of Josephson coupling in other van der Waals-based planar JJs have not been reported. We note that finite Josephson coupling in a vertical structure including a 5 to 10~nm-thick spacing layer of black phosphorus (bP) in between Nb slabs \cite{Chen2019,Xu2021} and in a vertical interconnect between MoS$_2$ and MoRe \cite{Ramezani2021} has been reported recently. In these structures, the semiconducting van der Waals material is used as a thin, almost transparent barrier, and the transport in the vertical direction, across the layers, is probed. There, however, gate tuning of the proximitized system is very difficult and can be achieved only laterally, resulting in a strong anisotropic modulation of carrier density in the semiconducting material \cite{Xu2021}. In this respect, planar junctions represent a scalable and versatile technology, where homogeneous gating can be easily obtained. In addition, planar junctions allow to investigate superconducting proximity effect in the extreme single layer (monolayer) limit, which is not possible in the vertical configuration.

In this work, we report for the first time on the fabrication and characterization of planar JJ devices based on few layer black phosphorus. bP is a layered semiconducting material, whose peculiar properties mainly stem from the anisotropic shape of its band structure, which affects both its electronic and optical properties \cite{Ling2015,Peruzzini2019,Keyes1953,Warschauer1963,Morita1986,Liu2014,CastellanosGomez2015,Telesio2020a}. Theoretical studies for superconducting JJs with monolayer bP have been put forward, with intriguing features that may emerge in ballistic samples in combination with ferromagnetic elements \cite{Linder2017,Alidoust2018,Alidoust2019}.

Here, we perform transport measurements on bP planar JJs with Nb contacts at cryogenic temperatures down to $T = 33$~mK, demonstrating, for the first time, supercurrent flow and clear evidence of Josephson coupling.


Exfoliated bP flakes were deposited on B-doped Si substrates. A geometry with four parallel contacts was designed to allow for the characterization of several JJs on a single BP flake. The devices were realized by electron beam lithography, with a procedure described in detail in the Methods section. Air exposure was minimized throughout all the fabrication process, and a combined oxygen-argon plasma process was optimized to enhance the interface quality. Superconducting Ti/Nb contacts of 10/60~nm thickness were deposited by sputtering technique. The critical temperature $T_c$ of the Ti/Nb electrodes was measured to be 8.4~K, which corresponds to a bulk superconducting gap of the electrodes of $\Delta_S = 1.76 k_B Tc = 1.3$~meV, in line with the expected value for bulk Nb \cite{Grosso2014,Guiducci2019,Guiducci2019a}.

\begin{figure}[t]
   \includegraphics[width=0.8\textwidth]{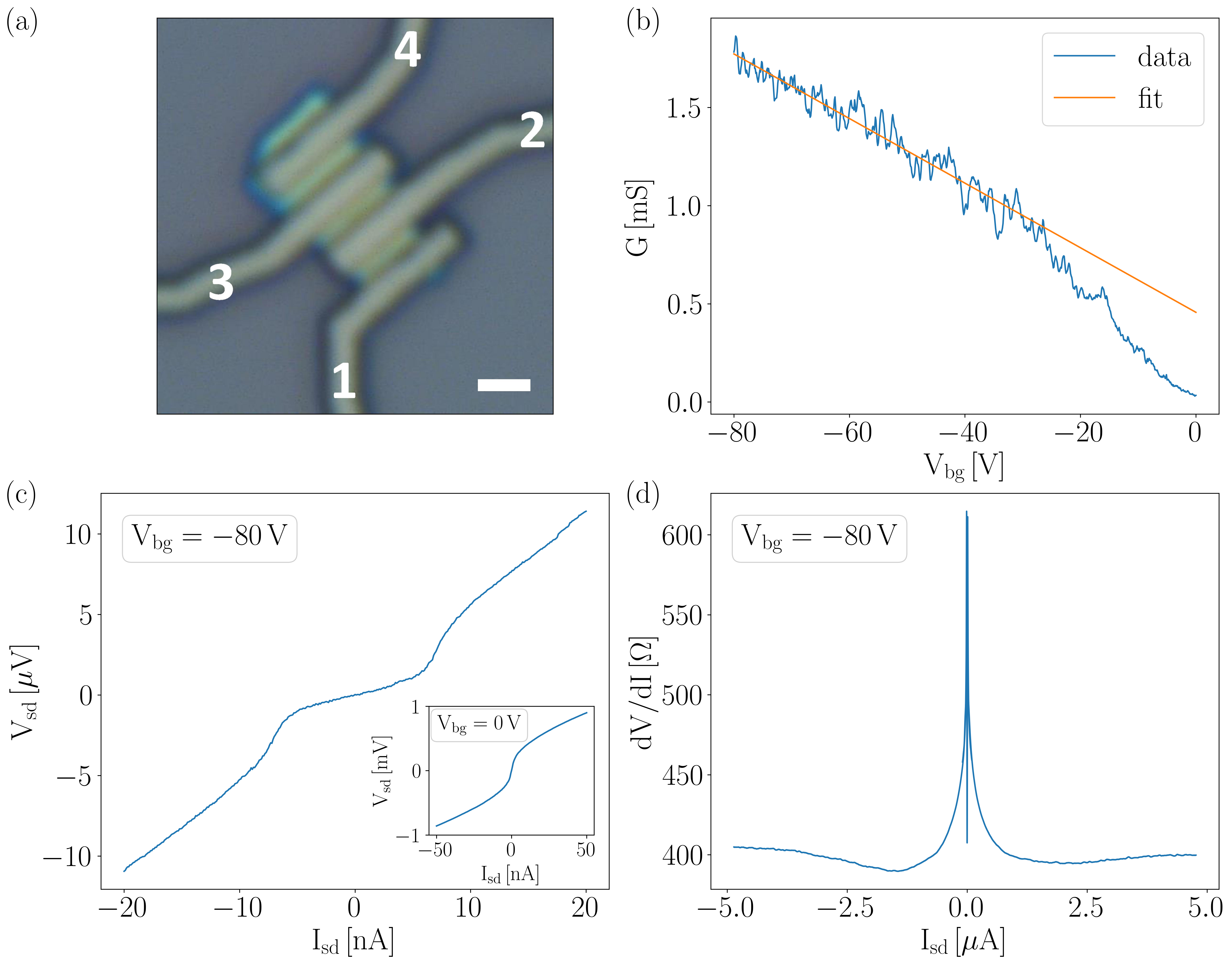}
   \caption{\label{fig1} (a) Optical microscopy image of the device. Contact numbers are indicated in the figure. Scale bar 1 $\mu$m. (b) Conductance $G$ versus back gate voltage $V_{bg}$ from $-80$~V to $0$~V, measured in current bias with a current of 20~nA. The straight line is a linear fit to the data from $V_{bg} = -80$~V to $-30$~V, used to calculate the hole mobility. (c) $V_{sd}$ as a function of $I_{sd}$ at a back gate voltage $V_{bg} = -80$~V. The inset shows the same for a back gate voltage of 0~V. (d) Differential resistance $dV/dI$ of the junction for a wide range of bias current values $I_{sd}$ at $V_{bg} = -80$~V. The central dip is due to the supercurrent in the junction. All data measured at $B = 0$~mT and $T = 33$~mK.}
\end{figure}

The devices shown in Fig.~\ref{fig1}(a) have length of the individual JJs $L \sim 500$~nm, while their widths $W$ range from 1.6~$\mu$m (between contacts $1-2$) to 2.0~$\mu$m (between contacts $2-3$ and $3-4$), due to a slight trapezoidal shape of the bP flake. Flake thickness of the few-layer bP, inferred from the optical microscopy image, is about 10~nm. In the following we will present results obtained between contacts 2 and 3. We underline that we have obtained consistent results from all devices.

Black phosphorus is an intrinsically p-type semiconductor \cite{Keyes1953,Warschauer1963,Morita1986}. For a back gate voltage $V_{bg} = -80$~V, the bP is deep in the p-type branch. Sweeping the back gate toward zero voltage, the Fermi level of the semiconductor is shifted into the band gap, which results in a strong increase in resistance, or, equivalently, a strong decrease in conductance. This is shown in the back gate sweep in Fig.~\ref{fig1}(b), which was measured in current bias with a current amplitude of 20~nA. Further increase of the back gate voltage to $+80$~V drives the device into the n-type regime, as shown in the Supplementary Information. From the back gate sweep, a field effect mobility $\mu = 360$~cm$^2$/(Vs) is obtained from a linear fit of $G$ vs. $V_{bg}$ in the range $-30$~V to $-80$~V.

Figure~\ref{fig1}(c) shows the $V-I$ characteristics of the device measured in a current bias configuration at a back gate voltage of $-80$~V. A clear and pronounced Josephson supercurrent with amplitude of 5~nA is observed. Once the bias current exceeds this critical value, the device switches from the superconducting to the normal state, with a resistance of $R_N = 560$~$\Omega$. As can be seen in Fig.~\ref{fig1}(c), the differential resistance around $I_{sd} = 0$ is not strictly zero but $R_0 \sim 200$~$\Omega$. This is partially due to the effect of thermal fluctuations in the junction and can be understood in the framework of the theory by Ambegaokar and Halperin \cite{Ambegaokar1969,Gross2005}. In fact, the McCumber parameter $\beta_c$ of the junction is $\beta_c = \omega_J \tau_{RC}$, with $\omega_J = 2 e V_c / \hbar$ the characteristic Josephson frequency ($\hbar$ is the reduced Planck constant, $V_c = I_c R_N$, with $I_c$ the critical current) and $\tau_{RC} = R_N C$ the decay time within the RCSJ model \cite{Gross2005,Belogolovskii2017,Zhukovskii2018}. Approximating the junction capacitance $C$ with a parallel plate capacitor model, we obtain $\beta_c = 1.2 \times 10^{-5} \ll 1$. Thus, the junction is in the overdamped regime, and the non-vanishing junction voltage even in the limit $I_{sd} \rightarrow 0$ can be attributed to a phase-slip resistance contribution. Consistent with this interpretation, the measured $V-I$ curves do not show any hysteresis, i.e., switching and retrapping current, $I_{sw}$ and $I_{rt}$, respectively, are equal.

Finally, we inspect the behavior of the junction at elevated bias, to look for any subharmonic gap features. Figure~\ref{fig1}(d) shows the differential resistance, measured at $V_{bg} = -80$~V. The supercurrent branch is resolved as the central dip, due to the reduced resistance in the $V-I$ curve at the origin. The differential resistance curve shown in Fig.~\ref{fig1}(d) does not show any evidence for multiple Andreev reflections (MARs), but instead a clear Schottky behavior, with an increased resistance at small bias. This indicates that at small bias the transparency of the S/N interface is not ideal, and the Josephson junction is better described as a SINIS structure, i.e., a superconductor-insulator-normal-insulator-superconductor hybrid system. From Fig.~\ref{fig1}(d), we can estimate the resistance $R_i$ of each S-N interface from the difference in differential resistance at low bias ($R_N = 560 \, \Omega = R_{bP} + 2R_i$) and at high bias ($R_{bP} = 400$~$\Omega$, see Fig.~\ref{fig1}(d)). This gives $R_i = 80$~$\Omega$. The hypothesis of opaque S/N interfaces is further supported by the lack of excess current as described in the Supporting Information.

The Schottky barrier is further increased at lower carrier concentration, as shown in the inset to Fig.~\ref{fig1}(c), which shows that at $V_{bg}=0$~V, in contrast to the observation at $V_{bg}=-80$~V, no supercurrent is observed, and the $V-I$ curves rather show a Schottky-like behavior. In line with the above discussion, this confirms the presence of a Schottky-barrier at the interface between bP and the superconductor, preventing a Josephson coupling at $V_{bg} = 0$~V. Also in the n-type regime, as shown in the Supplementary Information, the quality of the contacts does not improve, consistent with previous observations \cite{Telesio2020}, and the metal-semiconductor contacts are still dominated by the Schottky barrier, which prevents observation of a bipolar signal in supercurrent amplitude. Thus, in the following we will focus on the accumulation region.

From the data shown in Fig.~\ref{fig1}, important information on the transport in the junction can be extracted. The sheet resistance $R_s$ of bP at $V_{bg} = -80$~V is $R_s = (W/L) R_{bP} = 1.6$~k$\Omega/ \square$. Within a two-dimensional Drude model we obtain a carrier concentration {{at $V_{bg}=-80$~V}} of $n = 1.1 \times 10^{13}$~cm$^{-2}$. Using $\ell_e = \hbar k_F \mu /e$ and the Fermi wave vector $k_F = \sqrt{2\pi n}$ \cite{Ando1982}, we estimate the elastic mean free path to be $\ell_e = 20$~nm. This places the junction in the diffusive regime, since the mean free path is much smaller than the length of the junction, $\ell_e \ll L$. The coherence length of the N region (bP) is \cite{Chrestin1997} $\xi_N = \sqrt{ \hbar D / \left( 2 \pi k_B T \right)} = 290$~nm at 33~mK, which reduces to 100 nm at 300~mK. Here, $D = v_F \ell_e / 2 = \hbar^2 \pi / \left(e^2 m^* R_s \right)$ is the diffusion coefficient in the bP and $m^* = 0.41 m_0$ the average in-plane effective mass of holes in bP \cite{Qiao2014,Telesio2020a}. Therefore, $L$ is greater than $\xi_N$. This indicates that the device operates in the long-junction regime \cite{Mur1996,Grosso2014,Guiducci2019a}.

The natural energy scale for the proximity effect is the Thouless energy \cite{Courtois1996,Pannetier2000,Dubos2001} $E_{Th} = \hbar D / L^2$, here $E_{Th} = 6.0$~$\mu$eV. Thus, $\Delta_S \gg E_{Th}$. The thermal energy $k_B T$ is slightly smaller than $E_{Th}$ (here $k_B T = 2.8$~$\mu$eV) and also smaller than the Josephson energy \cite{Dubos2001} $E_{J} = \hbar I_c(T) / 2 e$, here $E_{J} = 10.3$~$\mu$eV, however, sufficiently large that thermal fluctuations would be relevant, supporting an interpretation of the data in terms of the Ambegaokar-Halperin picture \cite{Ambegaokar1969,Gross2005}.

\begin{figure}[pht]
   \centering
	 \includegraphics[width=0.5\textwidth]{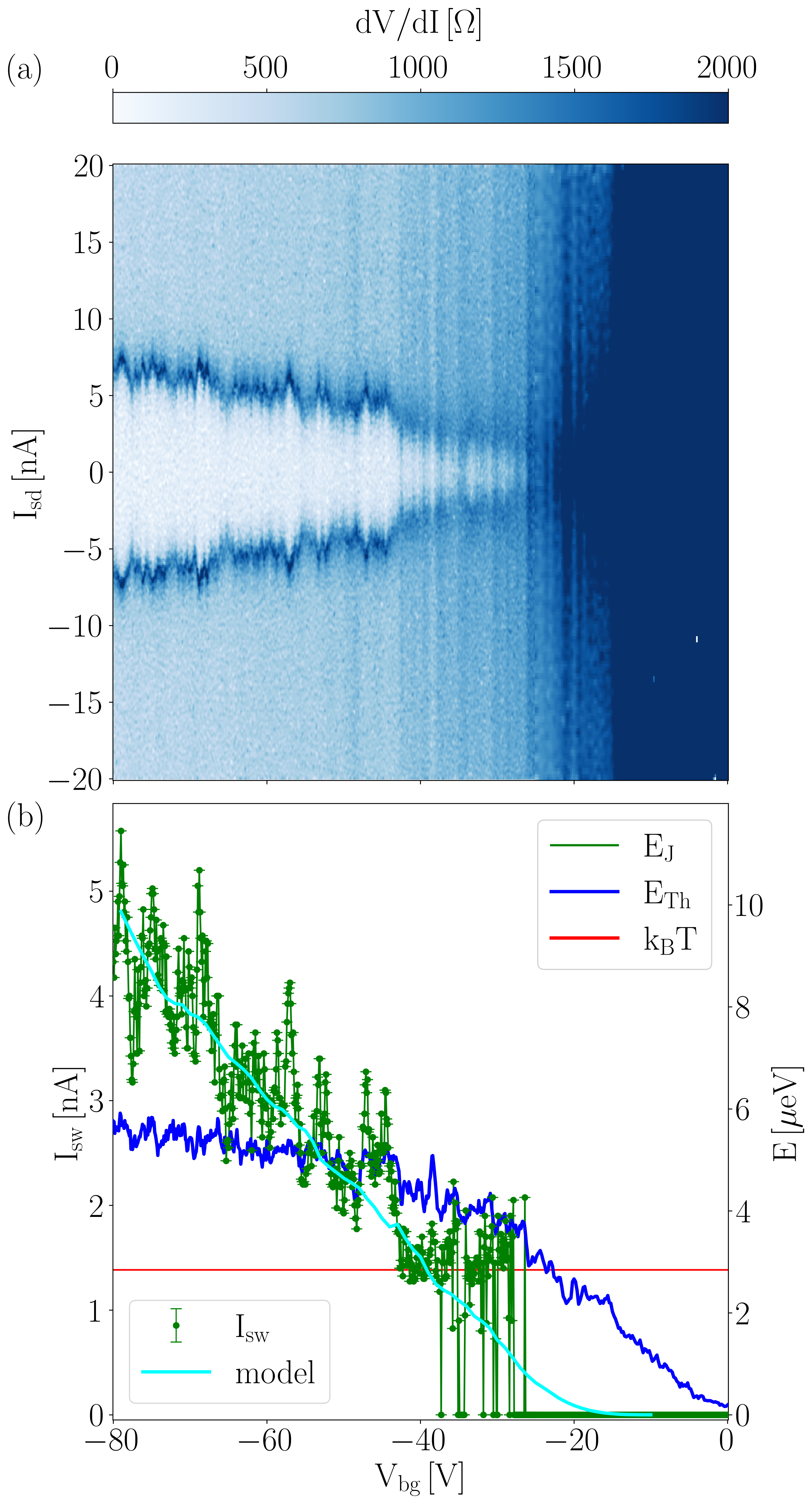}
   \caption{\label{fig2} (a) Differential resistance $dV/dI$ of the junction as a function of bias current $I_{sd}$ and back gate voltage $V_{bg}$. (b) Left axis: switching current $I_{sw}$, and right axis: the two relevant energy scales, Josephson energy $E_{J}$ and Thouless energy $E_{Th}$, as a function of back gate voltage $V_{bg}$, compared to the thermal energy $k_B T$. The two axes are linked via $E_{J} = \hbar I_{sw}(T) / 2 e$. A theoretical model for the critical current $I_c$ including a sizable interface resistance is also shown. The data in (b) were obtained from an analysis of the data in (a). $B = 0$~mT, $T = 33$~mK.}
\end{figure}

The supercurrent in the junction can be controlled by an external back gate. This is shown in Fig.~\ref{fig2}(a), which shows the differential resistance $dV/dI$ of the junction as a function of bias current $I_{sd}$ and back gate voltage $V_{bg}$. Reducing the back gate voltage from $-80$~V to $0$~V, the switching current is reduced from 5~nA to zero. This reduction in supercurrent is accompanied by an increase in the resistance of the normal branch (evaluated at $I_{sd} = 20$~nA). Figure~\ref{fig2}(b) shows, on the left axis, the switching current as a function of back gate voltage, while the right axis shows the corresponding energy scale, obtained via $E_{J} = \hbar I_{sw}(T) / 2 e$. Figure~\ref{fig2}(b) shows that a reduction of the back gate voltage reduces the Thouless energy $E_{Th}$ (via an increase in the sheet resistance). At about $V_{bg} = -40$~V, the Josephson energy $E_{J}$, which is proportional to the critical current, becomes smaller than the thermal energy $k_B T$. Beyond this point $I_{sw}$ displays strong thermal fluctuations, until a complete suppression is observed at $V_{bg} = -25$~V, a back gate voltage at which the Thouless energy becomes smaller than the thermal energy $k_B T$. Also at about $V_{bg} = -25$~V, the resistance at zero current $R_0$ becomes larger than the resistance in the normal branch, underlining the quenching of the Josephson coupling due to the increased Schottky barrier, as also shown in the $V-I$ curves in Fig.~\ref{fig1}(c).

The monotonic decay of $I_{sw} \left( V_{bg} \right)$ is well described in the framework of diffusive Josephson junctions with opaque S/N interfaces \cite{Kuprianov1988,Heikkilae2002,Tiira2017}. The details of the theoretical model are described in the Methods section. As shown in Fig.~\ref{fig2}(b), the model can accurately describe the monotonic decay of $I_{sw} \left( V_{bg} \right)$, adding further evidence for the presence of a Schottky-barrier at the interface between bP and the superconductor.

\begin{figure}[pth]
   \centering
   \includegraphics[width=0.5\textwidth]{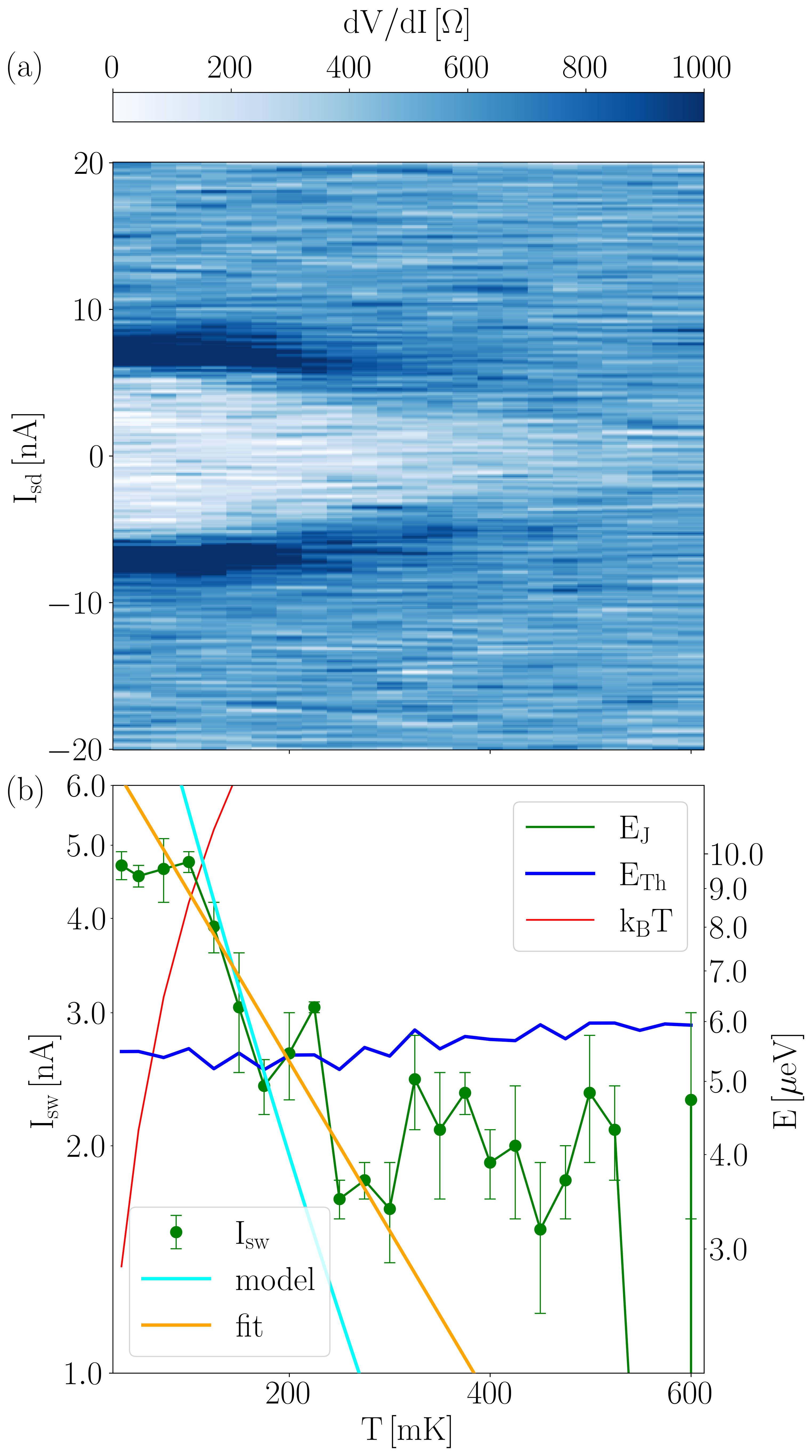}
   \caption{\label{fig3} (a) Differential resistance $dV/dI$ of the junction as a function of bias current $I_{sd}$ and temperature $T$. (b) Left axis: switching current $I_{sw}$ on a logarithmic scale as a function of temperature $T$. Error bars are the standard deviation of the measured data. The fitted line indicates an exponential decay of the switching current with temperature, which is also captured by the theoretical model. Right axis: the corresponding energy scale. $B = 0$~mT, $V_{bg} = -80$~V.}
\end{figure}

It is well known that an increase in temperature suppresses proximity-induced superconductivity. Figure~\ref{fig3}(a) reports the temperature-dependence of the differential resistance. The figure clearly shows that at $\sim 500$~mK, supercurrent is suppressed, and an Ohmic behavior of the junction is observed. Two effects lead to this evolution: first, the value of the switching current is reduced by temperature. At the same time, the slope $R_0$ of the $V-I$ curves at $I_{sd} = 0$ increases due to thermal broadening. Figure~\ref{fig3}(b) shows, on the left y-axis, the switching current as a function of temperature on a logarithmic scale. The right axis shows the corresponding energy scale. The graph shows that the Thouless energy $E_{Th}$ is approximately constant, since the normal resistance does not change significantly in this temperature window. At low temperatures, the supercurrent appears to be approximately constant until $k_B T$ becomes larger than $E_{J}$, possibly due to a saturation of electron temperature below 100~mK. Above that point, a linearly decreasing trend in switching current and thus in Josephson energy is observed (on a log-scale), compatible with an exponential decrease with temperature. However, when the Josephson energy becomes approximately equal to the Thouless energy, a deviation from this exponential decrease is observed, with $E_{J} \sim E_{Th}$, until finally at $T \sim 500$~mK the switching current drops to zero. An exponential decrease in switching current is typical of ballistic long junctions \cite{Mur1996,Guiducci2019a}, but a quasi-exponential temperature dependence of the critical current is also expected for a diffusive system in the long junction regime, as shown by Courtois and Sch\"on \cite{Courtois1995,Wilhelm1997,Dubos2001}. An analysis of the measured switching current with the function $I_{sw} = I_{0} \exp \left( -T/T^* \right)$ yields as best fit parameters $I_{0} = 7.3$~nA and $T^* = 194$~mK. The fit is also indicated in Fig.~\ref{fig3}(b). According to Wilhelm \textit{et al.}~\cite{Wilhelm1997}, $k_B T^* = 12 E^* / \pi$, with $E^* = E_{Th} \left( 1 + 0.7 R_i / R_{bP} \right)$ \cite{Heikkilae2002}, and thus from $T^*$ we get a value of the Thouless energy of 5.0~$\mu$eV, in excellent agreement with the value obtained from the data shown in Fig.~\ref{fig1}. Overall, this analysis underlines the importance of the Thouless energy as the relevant energy scale, since here $E_{Th} \ll \Delta_S$.

Also in this case, the temperature dependence of $I_{sw}$ can be well represented by a the theoretical model of a diffusive mesoscopic Josephson weak link with opaque interfaces (cyan curve in Fig.~\ref{fig3}(b)). Notably, the same model with clean interfaces yields an $I_c$ one order of magnitude higher, demonstrating the importance of the opaque interfaces in the estimation of $I_c$. 

\begin{figure}[t]
   \centering
   \includegraphics[width=\textwidth]{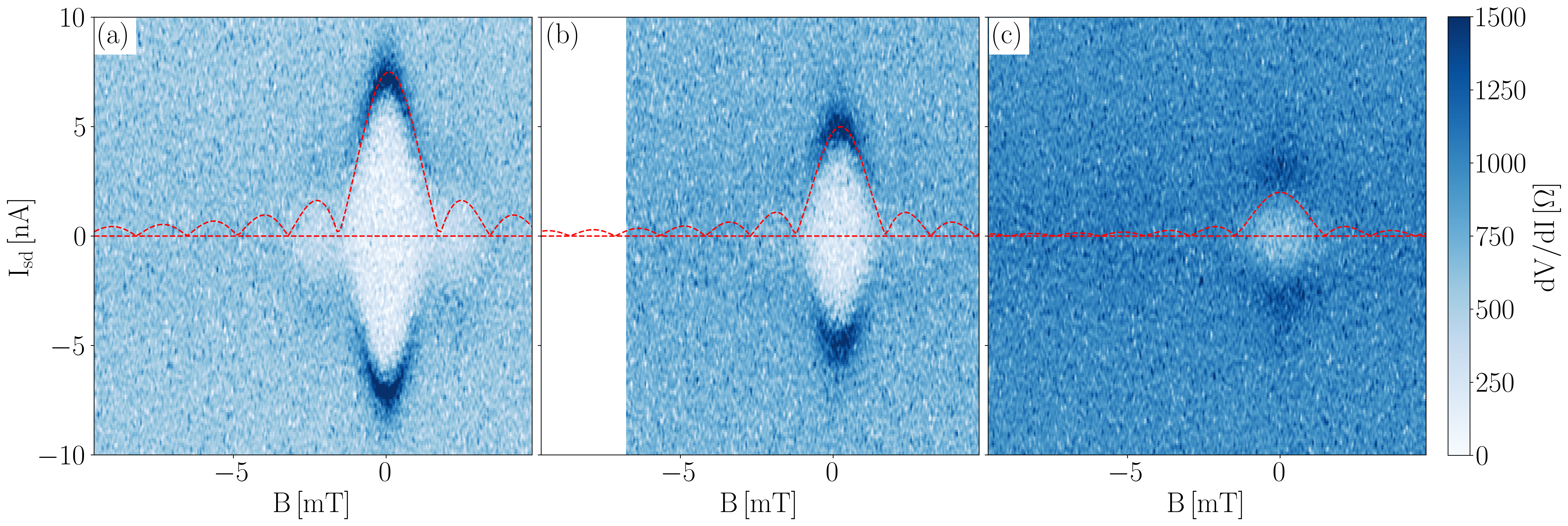}
   \caption{\label{fig4} Differential resistance $dV/dI$ of the junction as a function of bias current $I_{sd}$ and magnetic field $B$. (a) $V_{bg} = -80$~V, (b) $V_{bg} = -60$~V, (c) $V_{bg} = -40$~V. A fit to the standard Fraunhofer formula is plotted as dashed red line in each panel. $T = 33$~mK}
\end{figure}

A clear signature of a well established Josephson coupling in a JJ device is the presence of a Fraunhofer pattern, when a magnetic field perpendicular to the junction is applied. Figure~\ref{fig4} shows the differential resistance $dV/dI$ of the junction versus bias current $I_{sd}$ and perpendicular magnetic field $B$, measured for three different back gate voltages. The presence of a Fraunhofer pattern centered at $B = 0$ mT is clearly visible, consistent with a strong Josephson coupling at zero magnetic field. Applying a small magnetic field, the supercurrent is reduced, until for $\left| B \right| = 1.65$~mT it is zero, then it reappears, and at $\left| B \right| > 3.3$~mT it completely disappears. This behavior is characteristic of a Fraunhofer pattern with a central lobe and two side lobes, and the suppression of supercurrent by a small magnetic field clearly points at the proximity effect as the origin of the observed behavior of the $V-I$ curves around zero bias. The Fraunhofer patterns measured at different back gate voltages consistently follow the general trend discussed with Fig.~\ref{fig2}: the supercurrent is reduced by a lower negative voltage on the back gate, while the normal resistance is increased. However, the periodicity of the Fraunhofer pattern remains the same as for $V_{bg} = -80$~V, adding further support to the interpretation of these data in terms of induced superconducting proximity within the junction. 

These data were analyzed with the well-known conventional Fraunhofer formula $I_c \left( B \right) = I_c \left( 0 \right) \left| \sin \left(  \pi \Phi / \Phi_0 \right) / \left(  \pi \Phi / \Phi_0 \right) \right|$, with $I_c(0)$ the critical current at zero magnetic field, $\Phi$ the magnetic flux, and $\Phi_0 = h/2e$ the superconducting flux quantum.\cite{Vries2019,Baumgartner2021} The corresponding Fraunhofer patterns are included in Fig.~\ref{fig4} as dashed red lines. From the periodicity of the Fraunhofer pattern, we extract a characteristic area of $1.3$~$\mu$m$^2$, in good agreement with the geometrical area of the junction, $\sim 1$~$\mu$m$^2$. The slight difference between the two values can be explained by a magnetic field focusing due to the Meissner effect,\cite{Suominen2017,Monteiro2017,Guiducci2019a} or, alternatively, with an effective length of the junction of $L = 650$ nm. The actual geometry of the junction actually supports this latter interpretation, since the superconducting electrodes cover a considerable fraction of the bP flake.

In summary, we have presented transport measurements on bP planar JJs with Ti/Nb contacts performed at cryogenic temperatures down to $T = 33$~mK, demonstrating, for the first time, a sizable supercurrent up to 5~nA and unequivocal evidence of Josephson coupling. The supercurrent can be controlled by a back gate. Application of a small perpendicular magnetic field leads to the formation of a well-developed Fraunhofer pattern. Supercurrent signal disappears above a temperature of 500~mK. All these observations are clear manifestations of proximity-induced superconductivity in the bP planar junction and place bP in the spotlight as a valid 2D van der Waals material for applications in quantum technology. We anticipate that optimization of the Ohmic contacts to the bP will improve device performance and eventually allow to observe ambipolar supercurrent in such devices. Further progress in device fabrication, e.g.~by encapsulation of the bP flakes in hexagonal Boron Nitride, will allow to realize ballistic junctions, in which the crystalline anisotropy of bP might be employed for innovative devices and quantum sensors. Among all, bP devices are exploited as widely tunable infrared photodetectors \cite{Viti2015,Guo2016,Giordano2018}, and optimized bP-based planar Josephson junctions are thus a promising platform for novel infrared single photon detectors \cite{Walsh2017,Walsh2021}.

\section{Methods}

\subsection{Device fabrication}

BP crystals were fabricated following a well-established procedure \cite{NILGES2008}. For details, see the Supporting Information. There we also show a Raman spectrum of the flake used in this work. The high reactivity of exfoliated bP represents the major challenge for the realization of high-quality devices, including JJ-like devices, where the quality of the interfaces represents one of the main bottlenecks. To address this issue, bP exfoliation was carried out in a glove bag under a  nitrogen atmosphere. After the transfer of the flakes onto the Si wafer substrate, the samples were immediately coated with a layer of protective polymer (poly methyl-methacrylate, PMMA), which is also the lithographic resist. The geometry of the device was defined by electron beam lithography (EBL) using a Zeiss Ultraplus SEM equipped with Raith Elphy Multibeam software. After development of the resist, a combined and optimized cleaning procedure was applied to the sample \cite{Cappelli2020}. First, a mild oxygen plasma of 10 W for 1 minute with 40 SCCM of oxygen was performed to efficiently remove all the resist residuals. Then the sample was transferred to the vacuum chamber for the metals sputtering, and an \textit{in-situ} cleaning with Ar plasma was performed at 50 mtorr with 4 W power for 30 minutes. This last cleaning step was performed after the pre-sputtering of the two metallic targets for metal deposition, during which the sample was preserved from contamination thanks to a closed shutter and a rotating carousel that allowed to move the sample away from the plasma source. After 10~nm Ti and 60~nm of Nb sputter deposition, the sample underwent a fast lift-off in acetone at 55 $^\circ$C. Then it was immediately coated with a bilayer of methyl methacrylate methacrylic acid copolymer and PMMA, to guarantee protection from oxidation. This protection layer is more than 500~nm thick, therefore holes for the bonding pads have been opened in a second EBL step.

\subsection{Transport measurements}

The low-temperature transport data were measured in a filtered closed-cycle dilution refrigerator from Leiden Cryogenics with a base temperature of $T = 33$~mK. $V-I$ curves were measured in DC current biasing, using a Yokogawa DC source on a 10~M$\Omega$ bias resistor. The voltage drop across the junction was measured in four-probe configuration with a Delta voltage preamplifier (gain $10^4$) and an Agilent multimeter, while the current was measured with a Delta current preamplifier (gain $10^7$) and another Agilent multimeter. The back gate was biased with a Keithley DC source. We have verified that the loss current from gate to sample was below the detection limit of the source meter. The differential resistance measurements shown in Fig.~\ref{fig1}(d) were obtained with a DC + AC signal, using standard AC lock-in technique with current excitation in the range 5~nA to 10~nA.

\subsection{Model}

The observed behavior of the critical current $I_c$ of the junction can be modeled by solving the associated quasi-classical Usadel equation\cite{Usadel1970}. To properly reproduce the experimental data, a sizable S/N interface resistance ($R_i$) has been introduced, considering a diffusive Josephson Junctions with opaque S/N interfaces. In this case, the critical current can be obtained by \cite{Kuprianov1988,Heikkilae2002,Tiira2017}:
\begin{equation}
   \label{eq:dirtyNum}
   I_c = \frac{2 \pi k_B T}{e R_{bP}}  \sum_{n=0}^{\infty}  \frac{\left( k_{\omega_n} L \right) \Delta^2}{\left( \omega_n^2 + \Delta^2 \right) \left[ \alpha \sinh{\left( k_{\omega_n} L \right)} + \beta \cosh{ \left( k_{\omega_n} L \right)} \right]},
\end{equation}
where the sum over Matsubara frequencies $\omega_n = (2n+1) \pi k_B T$ is carried out numerically. Here, $k_B$ is the Boltzmann constant, $k_{\omega_n} = \sqrt{2 \omega_n / \left( \hbar D \right)}$ , $R_{bP} = R_N - 2 R_i$ is the resistance of the junction excluding the interfaces, $\alpha = 1 + \left( r k_{\omega_n} L \right)^2$, $\beta = 2 r k_{\omega_n} L$, and $r = R_i / R_{bP}$. Using Eq.~\ref{eq:dirtyNum} it is possible to calculate the temperature dependence of the critical current (Fig.~\ref{fig3}(b)). The same model has been used to simulate the evolution of the critical current as a function of the gate voltage (Fig.~\ref{fig2}(b)). The gate affects the resistance of the junction measured at low bias $R_N(V_{bg})$ and as a consequence the diffusion constant $D \propto 1 / R_N(V_{bg})$. The best fit to the experimental data was obtained with $L = 650$ nm and $T = 110$~mK. The variation of $r$ with back gate voltage was approximated by a linear function $r = 3.86 \times 10^{-3}$ V$^{-1} \cdot V_{bg} + 4.83 \times 10^{-1}$, which was the best fit to a series of $V-I$ curves measured in a wide bias range at different back gate voltages.

\begin{acknowledgement}

The authors thank Stefano Roddaro for granting access to a 2.4 K cryostat in the initial phase of this research, Filippo Fabbri for measuring the Raman spectrum, and Sebastian Bergeret per useful discussions. We acknowledge the European Research Council for funding the project PHOSFUN {\it Phosphorene functionalization: a new platform for advanced multifunctional materials} (Grant Agreement No. 670173) through an ERC Advanced Grant to MP. ES and FG acknowledge the European Research Council under Grant Agreement No. 899315 (TERASEC), and  the  EU’s  Horizon 2020 research and innovation program under Grant Agreement No. 964398 (SUPERGATE) and No. 800923 (SUPERTED) for partial financial support.

\end{acknowledgement}

\begin{suppinfo}

The Supporting Information is available free of charge.

\begin{itemize}
  \item Additional information on device fabrication, Raman spectroscopy data, and additional transport characterization.
\end{itemize}

\end{suppinfo}

\bibliography{Bibliography-2021-07-22}

\end{document}